\documentclass[epj]{svjour}
\usepackage{graphicx}
\begin{document}
\title{Enhancement of octupole strength in near spherical nuclei}
\author{L. M. Robledo\inst{1} 
\thanks{luis.robledo@uam.es}%
}                     
%
%
\institute{Dep. F\'\i sica Te\'orica, Facultad de Ciencias, Universidad Aut\'onoma de Madrid, E-28049 Madrid, Spain
}
\date{Received: date / Revised version: date}
%
\abstract{The validity of the rotational formula used to compute E1 and 
E3 transition strengths in even-even nuclei is analyzed within the 
Generator Coordinate Method framework based on mean field wave 
functions. It turns out that those nuclei with spherical or near 
spherical shapes the E1 and E3 strengths computed with this formula are 
strongly underestimated and a sound evaluation of them requires angular 
momentum projected wave functions. Results for several isotopic chains 
with proton number equal or near magic numbers are analyzed and 
compared with experimental data. The use of angular momentum projected 
wave functions greatly improves the agreement with the scarce 
experimental data.
} 
\maketitle
\section{Introduction}
\label{intro}

The study of octupole correlations is of fundamental importance to 
understand the properties of low lying collective states of negative 
parity in atomic nuclei \cite{RevModPhys.68.349}. Octupole deformation 
can be static, when the nuclear wave function breaks the symmetry under 
spatial reflection leading to a non-zero octupole moment or dynamic 
when it preserves the symmetry but quantum fluctuations involving 
octupole shapes are relevant \cite{Gaffney2013}. The observables of 
interest associated with octupole correlations include the excitation 
energy of collective negative parity states and the electromagnetic 
transition strengths of E1 and E3 type connecting them to the ground 
state. Although the presence of permanent octupole deformation is 
scarce in even-even nuclei dynamical octupole correlations are 
pervasive over all the periodic table. In spherical nuclei octupole 
vibrational excited states are also common and they are characterized 
by very collective E3 transition strengths.  

The study of the properties of collective negative parity states in 
even-even nuclei has been recently addressed using several variants of 
the Gogny force in the framework of the Generator Coordinate Method 
(GCM) \cite{PhysRevC.84.054302} with the axial octupole moment as 
generating coordinate. Results for the excitation energy of the 
collective negative parity state as well as the E1 and E3 transition 
strengths to the ground state were obtained. The comparison with 
experimental data revealed a systematic over-estimation of the 
excitation energy. Concerning the transition strengths, there was a 
systematic behavior indicating a much better reproduction of the 
strength in well deformed nuclei.

In addition to this study, the impact of static and dynamic octupole 
correlation on the binding energies of the ground state of even-even 
nuclei has been considered in \cite{0954-3899-42-5-055109}. It was 
shown there that all even-even nuclei show octupole dynamic 
correlations with energy gains in the range in between one and two MeV 
but with a smooth behavior as a function of proton or neutron number.

A subsequent analysis of the E3 transition strengths obtained with 
mean field states, revealed \cite{PhysRevC.86.054306}
that the rotational formula used to compute them is not valid in spherical
or weakly quadrupole deformed systems. A careful analysis of the general
structure of the wave function near sphericity revealed that, in that case,
the transition strength had to be up to a factor 7 ($2I+1$ with $I=3$) larger
than the value provided by the rotational formula. 

\section{Theoretical framework}
\label{sec:Theo}

As it is customary in nuclear structure, the theoretical framework is 
based on the mean field Hartree- Fock- Bogoliubov (HFB) method 
\cite{RS80}. A set of HFB intrinsic wave functions is generated using 
the axially symmetric octupole moment $Q_{30}$ as a constraint and 
considering a mesh of octupole moments adapted to the problem at hand. 
The wave functions generated this way break reflection symmetry. In 
many cases, they also break rotational invariance and particle number 
and therefore the wave functions belong to the category of intrinsic 
wave functions, that is the symmetry quantum numbers have to be 
restored in order to compute quantities in the laboratory frame. For 
the energy density functional (EDF) we have used the Gogny EDF 
\cite{DechargeGogny1980} with the D1S parametrization\cite{berger1984}. 
Most of the quantities discussed in the present paper have also been 
obtained with the D1M \cite{goriely2009}  and D1N \cite{chappert2008}. 
The comparison with the D1S results reveal many similitudes and 
therefore only the latter have been considered here.  A highly 
optimized computer code using second order gradient information 
\cite{PhysRevC.84.014312} is used in the calculations. In the computer 
code, the quasiparticle operators are expanded in a harmonic oscillator 
basis containing a number of major shells that depends on the mass 
number of the nucleus to be treated and large enough as to guarantee 
the convergence of excitation energies and transition strengths - see 
\cite{PhysRevC.84.054302} for details. The set of HFB wave functions 
$|\varphi (Q_{3})\rangle$ is subsequently used in several beyond mean 
field schemes discussed below.

\subsection{Parity Projection}
\label{subsec:1.1}

The first step beyond the mean field is to recover some of the broken 
symmetries of the intrinsic states. To restore reflection symmetry 
(associated to the parity quantum number) we use the parity projection 
operator $P_{\pi}=1+\pi\hat{\Pi}$ which is a linear combination of the 
symmetry operations: identity and parity $\hat{\Pi}$. The relative 
amplitudes of the linear combination determine the quantum number 
$\pi=\pm 1$ of the restored symmetry. Applying the projector to the 
intrinsic states $|\varphi (Q_{3})\rangle$ we obtain states with good 
parity $\pi$ that can be used to compute observable quantities like the 
energy. As there are two possible values of the parity, there are also 
two energies $E_{\pi}$. In the restricted variation after projection 
(RVAP) scheme \cite{RS80} the parity projected energies $E_{\pi}$ are 
computed for each member of the set of intrinsic states $|\varphi 
(Q_{3})\rangle$ and the two minima determine the optimal intrinsic 
states for each parity. In even-even nuclei where the positive parity 
state is always more bound than the negative parity one, the lowest 
positive parity state is associated to the ground state whereas the 
negative parity one is associated to the lowest lying negative parity 
state. This negative parity state that can be, depending on the 
deformation of the nucleus, the band head of a negative parity 
rotational band in the case of a quadrupole deformed nucleus or a 
$3^{-}$ vibrational state in case of a weakly or spherical nucleus. In 
order to compute transition strengths between the negative parity and 
the ground state the rotational formula is assumed. For instance, 
assuming axially symmetric states
\begin{equation}
\label{be3def} 
 B(E3, 3^-\rightarrow 0^+) = \frac{e^2}{4\pi}  \langle \Psi_{-} | \hat Q_{30}\frac{1+t_z}{2} |\Psi_{+}\rangle^2. 	
\end{equation}
with $\hat Q_{30}=z(z^2-\frac{3}{2}(x^2+y^2))$ the $K=0$ component of 
the octupole moment. The rotational formula is based on the rotational 
assumption and is used to connect intrinsic quantities with quantities 
in the lab system like transition strengths. This scheme is used in 
\cite{PhysRevC.84.054302} to compute E1 and E3 transition strengths of 
essentially all relevant even-even nuclei. A comparison with the E3 
experimental compilation of Ref \cite{Kibedi2002} reveals that the 
theoretical predictions agree much better with experiment when the 
nucleus is well quadrupole deformed.

\subsection{Generator coordinate method}
\label{subsec:1.2}

The generator coordinate method (GCM) considers linear combinations of
a set of relevant mean field states (like the set $|\varphi (Q_{3})\rangle$ 
mentioned above) to create correlated wave functions
\begin{equation}
	|\Psi_{\sigma}\rangle = \int \, dQ_{3} f_{\sigma} (Q_{3}) |\varphi (Q_{3})\rangle
\end{equation}
labeled by $\sigma$. Usually, instead of continuum variables $Q_3$ a 
discrete set $\{Q_{3,\: i},i=1,\ldots,N\}$ is considered; the integral 
is then replace by a sum. The amplitudes $f_{\sigma}(Q_3)$ are 
determined by the variational principle on the energy what leads to the 
Hill-Wheeler equation
\begin{equation}
	\int d Q_{3}' \langle \varphi(Q_{3})|(\hat{H}-E_{\sigma})|\varphi(Q_{3}')\rangle f_{\sigma} (Q_{3}')=0 
\end{equation}
requiring the norm 
$$\mathcal{N} (Q_{3}, Q_{3}')=\langle 
\varphi(Q_{3})|\varphi(Q_{3}')\rangle
$$ 
and Hamiltonian 
$$
\mathcal{H} 
(Q_{3}, Q_{3}')=\langle \varphi(Q_{3})|\hat{H}|\varphi(Q_{3}')\rangle
$$
overlaps. These quantities are computed with the generalized Wick 
theorem and the pfaffian formula 
\cite{Balian.69,Rob94,PhysRevC.79.021302,PhysRevLett.108.042505}. The 
evaluation of the hamiltonian overlap for density dependent 
interactions is still subject to some conceptual difficulties - see 
\cite{Robledo10} for a discussion involving the octupole degree of 
freedom. Once the $f_{\sigma} (Q_3)$ amplitudes are obtained, the 
transition strength is computed with the formula of Eq \ref{be3def} but 
using for the $|\Psi_+\rangle$ and $|\Psi_-\rangle$ wave functions the 
GCM states obtained for the ground and first excited state, 
respectively. As in the RVAP parity projection calculation discussed 
above, the results obtained for a compilation of even-even nuclei 
\cite{PhysRevC.84.054302} and the comparison with experimental data 
revealed a pattern associated with the quadrupole deformation of the 
nucleus: when the nucleus posses a weak quadrupole deformation (measured
by the experimental ratio $E_{4^+}/E_{2^+}$) the 
agreement is worse than when the nucleus is well deformed. This 
patterns seems to indicate that the rotational formula, obtained under 
the assumption of a strongly deformed system \cite{bohr1975,RS80}, is 
not valid when the nuclear deformation falls below certain limit that 
depends on mass number. Therefore, a sound calculation of transition 
strengths requires going beyond the rotational formula and therefore 
requires the introduction of angular momentum projected (AMP) wave functions.

\subsection{Angular momentum projection}
\label{subsec:1.3}

Angular momentum projection is the last step in our beyond mean field 
scheme but it will only be used here to compute transition strengths. 
The reason is that the calculation of transition strengths only 
requires the evaluation of one body operator overlaps which is a 
relatively inexpensive computational task. This is in contrast with the 
evaluation of two body operators, like in the evaluation of hamiltonian 
overlaps, required to fully implement an angular momentum projected GCM 
where the $f_\sigma$ amplitudes are determined by a projected 
Hill-Wheeler equation \cite{RodGuz02}. In the next section, we will 
project onto good angular momentum  the GCM wave functions of Sect 
\ref{subsec:1.2} without taking into account the effects of AM 
projection in the $f_{\sigma}$ amplitudes. The projected wave functions 
will be used to compute norms and transition strengths. 

In the present calculation an important simplification comes from the 
restriction to axially symmetric intrinsic states. Due to this symmetry
the angular momentum projected state only involves $K=0$ intrinsic states and
is written as
\begin{equation}
	|(\Psi_{\sigma})^{J}_{M}\rangle = N^{J}_{M} P^{J}_{M0} |\Psi_{\sigma}\rangle
\end{equation}
where $N^{J}_{M}$ is a normalization constant and $P^{J}_{MK}$ is the 
standard angular momentum operator 
\begin{eqnarray}
	P^{J}_{MK} & = & \frac{2J+1}{8\pi^2} \int d\alpha d\beta d\gamma \mathcal{D}^{J\,*}_{MK} (\alpha,\beta,\gamma) \\ \nonumber
	& \times & \exp(\frac{-i}{\hbar}\alpha J_{z}) \exp(\frac{-i}{\hbar}\beta J_{y}) \exp(\frac{-i}{\hbar}\gamma J_{z})
\end{eqnarray}
To compute the transition strengths the calculation of the overlap of
the electric multipole moment operator $Q_{\lambda \mu}$ (which is a tensor of
rank $\lambda$) between 
projected states is required
\begin{equation}
	\langle (\Psi_{\sigma'})^{J'}_{M'}|Q_{\lambda \mu}|(\Psi_{\sigma})^{J}_{M}\rangle
\end{equation}
For the E1 strengths the standard dipole operator, that takes into 
account center of mass effects, is considered instead. This is a 
quantity that can be evaluated using the magic of the Wigner-Eckart 
theorem and the final expression for the $B(E\lambda)$ only involves 
reduced matrix elements which are given in the present framework, 
restricted to axially symmetric intrinsic states,in terms of an 
integral in the rotation angle $\beta$ -- see \cite{RodGuz02} for a 
detailed derivation. Another characteristic of the calculation of 
even-even nuclei is that the intrinsic states are, by construction, 
eigenstates of the simplex operator $\mathcal{S}=\Pi R_{y}(\pi)$. As a 
consequence of this discrete symmetry, the projected states with 
angular momentum $J$ must necessarily have the ``natural parity" 
$\pi=(-1)^{J}$ and therefore only the $0^+$, $1^-$, $2^+$, $3^-$, etc 
are accessible.

A similar calculation aimed at the evaluation of E1 and E3 strengths 
was carried out in \cite{PhysRevC.86.054306} but using instead of the 
GCM correlated wave functions the intrinsic wave functions obtained 
with the parity projected procedure of Sec \ref{subsec:1.1}. The 
differences between  the results obtained by projecting the parity 
projected RVAP intrinsic states and the GCM ones were first analyzed in 
\cite{2014EPJWC..6602091R} where a few selected examples were 
considered in detail. The results discussed in this reference are 
summarized in Table \ref{tab:Various} where an assorted set of nuclei 
regarding their ground state quadrupole deformation is discussed. Two 
facts stand out in this table: first the strong impact in the E3 
transition strength of using AM projected wave functions when the 
nucleus is spherical as in the Pb case where we go from the 7.7 
Weisskopf units (W.u) given by the rotational formula to 22 W.u. when 
AM projected wave functions are considered. The second fact is the 
relevance of using GCM wave functions instead of the parity RVAP for 
the calculation of those strengths as in several examples like 
$^{64}$Zn, $^{158}$Gd or even $^{224}$Ra there are noticeable 
differences representing up to a 20\% change. The reason is that in 
those nuclei, the quadrupole deformation shows relevant variations 
along the set of intrinsic wave functions $|\varphi(Q_3)\rangle$. Those 
variations can not be accounted for by just considering two intrinsic 
wave functions as in the RVAP case. An extreme example is $^{64}$Zn 
that requires the explicit treatment of quadrupole and octupole 
degrees for freedom to obtain reasonable results. The theoretical 
predictions for the excitation energy of negative parity states obtained 
without AMP are given in the supplemental material of \cite{PhysRevC.84.054302}.
The corresponding AMP are much more computationally intensive and they
are not available yet. Exploratory results reveal that the excitation
energies do not differ much in the two cases, probably as a consequence
of the scalar nature of the Hamiltonian, much simpler than the tensor
one of the transition operators.

\begin{table}
\centering
\caption{\protect{$B(E3,3^-_1 \rightarrow 0^+_1)$} transition strengths in W.u. 
for the different approaches considered. 
See text for notation.
In all the cases, the suffix AMP represents Angular Momentum Projected calculations. The
last column is the quadrupole deformation parameter  $\beta_{2}$ for the ground state.
Experimental data taken from \cite{Gaffney2013} and \cite{Kibedi2002}.
\label{tab:Various}}
\begin{tabular}{lcccccc}
\hline
 & Exp & PP      & PP          & GCM    & GCM        & $\beta_{2}$      \\ 
 &     &    RVAP &         AMP &  Q3    &    AMP     &    (GS)          \\ \noalign{\smallskip}\hline\noalign{\smallskip}
$^{20}$Ne & 13  & 9.73 & 12.78 & 12.46 & 12.31 & 0.66 \\
$^{64}$Zn & 20 $\pm$ 3 & 10$^{-4}$ & 0.6 & 2.83 & 5.75 & -0.21 \\
$^{158}$Gd & 12  & 9.43 & 10.29 & 13.73 & 12.64 & 0.34 \\
$^{208}$Pb & 34  & 7.07 & 23.06 & 7.67 & 21.93 & 0 \\
$^{224}$Ra & 42 $\pm$ 3 & 71.39 & 71.39 & 70.16 & 61.51 & 0.18 \\\noalign{\smallskip}\hline\noalign{\smallskip}
\end{tabular}
\end{table}

An interesting quantity is the probability of finding a state with angular
momentum $J$ and parity $\pi$ in the intrinsic GCM wave function $|\Psi_\sigma\rangle$
and given by
\begin{equation}
\mathcal{P}^{J,\pi} = \left| \langle \Psi_\sigma | \hat{P}^{J,\pi} |\Psi_\sigma\rangle\right|^2
\label{eq:PJ}
\end{equation}
where $\hat{P}^{J, \pi} $ is the standard angular momentum projector for axially symmetric
configurations defined previously. In the present calculation, and due to
the simplex symmetry of the intrinsic states, the only relevant probabilities
are those for the states fulfilling the natural parity rule $\pi=(-1)^J$. For the
other states not fulfilling this rule the probabilities are exactly zero. For well deformed 
intrinsic states the 
probability distribution as a function of $J$ is expected to be broad with small amplitudes for
a large range of $J$ values \cite{RS80}. On the other hand, when dealing with 
nearly spherical ground states the angular momentum $J=3$ of the ``octupole
phonon" couples to the $J=0$ of the ground state and therefore a probability
distribution $\mathcal{P}^{J,\pi}$ strongly peaked at $J^\pi=3^-$ is to be 
expected for the first negative parity state. This will be illustrated in the 
section devoted to the results below. 


\section{Results}
\label{sec:Results}%

\begin{figure}
\resizebox{\columnwidth}{!}{%
  \includegraphics{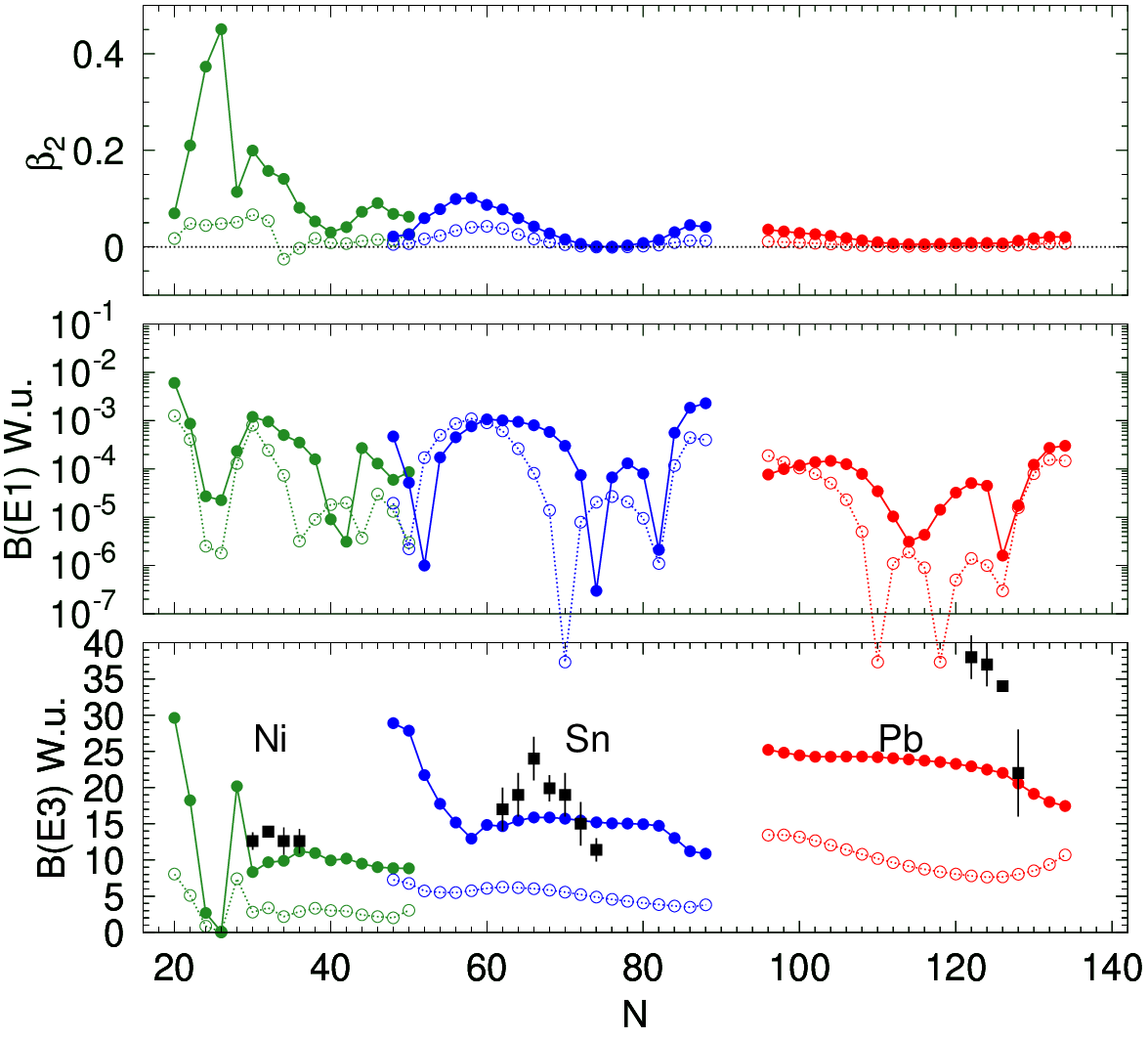}
}
\caption{Lower panel: the $B(E3,3^-\rightarrow0^+)$ transition
strength (in W.u.) are plotted as a function of the neutron number of 
each isotope. Three isotopic chains corresponding to the Ni ($Z=28$), 
Sn ($Z=50$) and Pb ($Z=82$) are plotted in green, blue and red, respectively.
Open symbols connected with dotted lines represent the results obtained 
with the rotational formula whereas full symbols connected with full lines
correspond to the calculation using angular momentum projected wave functions.
Disconnected full squares with error bars correspond to the experimental
results taken from the compilation of Ref \cite{Kibedi2002}.
Middle panel: $B(E1,1^-\rightarrow0^+)$ transition
strength (in W.u.) are plotted as a function of the neutron number with the
same color scheme as in the previous panel. Upper panel: the average $\beta_2$
deformation parameter for the $0 ^+$ ground state (open symbols) and the
lowest negative parity state (full symbols) are plotted as a function of
neutron number.}
\label{fig:1}       
\end{figure}

The purpose of this section is to present and discuss the results
obtained in the one dimensional GCM framework, using the octupole
moment $Q_{30}$ as generating coordinate, and computing the transition
strengths using the AMP wave functions, as described in Sect.~\ref{sec:Theo}.
The nuclei considered in this study are those susceptible to show a
significant failure of the rotational formula to compute transition 
strengths and include isotopic chains of nuclei with a proton number
in the neighborhood of a magic number. We will consider isotopic
chains of the elements Fe, Ni and Zn corresponding to the magic proton
number 28, the chains of  Cd, Sn, and Te  for the magic number Z=50 and
finally the chains of  Hg, Pb, and Po corresponding to the magic 
proton number of 82. 

In Fig \ref{fig:1} we present the main results of the present study for
the semi-magic isotopes of Ni, Sn and Pb plotted as a function of 
neutron number. In the upper panel, the average quadrupole deformation
parameter $\beta_{2\, \mathrm{av}}$ for the intrinsic GCM ground state and the 
intrinsic first negative parity excited state 
\begin{equation}
	\beta_{2\, \mathrm{av}} (\sigma) = \sqrt{5/(16\pi)} 4 \pi / (3 r_0^2 A^{5/3} ) \langle \Psi_{\sigma} | \hat{Q}_{20} | \Psi_{\sigma} \rangle
\end{equation}
is plotted. In the Pb chain both quadrupole deformation parameters are
very small, corresponding to spherical and negative parity excited states.
In the Sn chain a similar behavior is observed for isotopes with large
neutron number, close to $N=82$. However for lower neutron numbers the negative parity 
excited state acquires a deformation of around 0.1 that goes back to 
spherical at $N=50$. In the Ni case, the situation is a bit different, 
with spherical ground states near $N=50$, but for the negative parity
state there is some deformation even in this case. At lower neutron
numbers, close to $N=28$, the ground state is slightly deformed and the
excited state is strongly deformed. 

In the lower panel of Fig \ref{fig:1}
the $B(E3,3^{-}\rightarrow 0^{+})$ transition strengths in W.u. are plotted
for the three isotopic chains. Open symbols represent the results obtained
from the intrinsic moments using the rotational formula. Full symbols are
used to represent the results obtained with the AM projected states. In 
the Pb case, where both the $0^{+}$ ground state and the first negative
parity excited states are nearly spherical the projected results are in
the whole isotopic chain a factor nearly three larger than the rotational
formula results. The AM projected results are much closer to experimental
data than the rotational formula results although there  some
noticeable differences still remain. In the Sn case, where both states
are nearly spherical for most of the neutron number values, the enhancement
due to projection is also strong and in the range of a factor three. There
is an exception near N=50 where the negative parity excited state is
also spherical and the enhancement factor grows up to nearly five. It would
be interesting to measure the E3 strength in $^{100}$Sn to asses this effect.
The comparison with experimental data is very satisfactory when the AM projected
wave functions are used, clearly indicating the necessity of using this kind
of wave functions to compute transition strengths. Finally, in the Ni case, 
the trend is similar with strong enhancements of the E3 strengths when both
the ground state and the negative parity one are near spherical and some
accidents like in $^{54}$Ni where the negative parity state is strongly
deformed and the E3 strength shows a pronounced dip. As in the Sn and Pb
cases, the agreement with experiment is very good when the AM projected
wave functions are used in the calculation. It would be interesting to
measure the E3 strength in both $^{54}$Ni and $^{56}$Ni to see if the 
predicted strong variation is observed experimentally as it would serve
as an indicator of the strong quadrupole deformation of the excited state in 
$^{54}$Ni.

%
\begin{figure}
\resizebox{\columnwidth}{!}{%
  \includegraphics{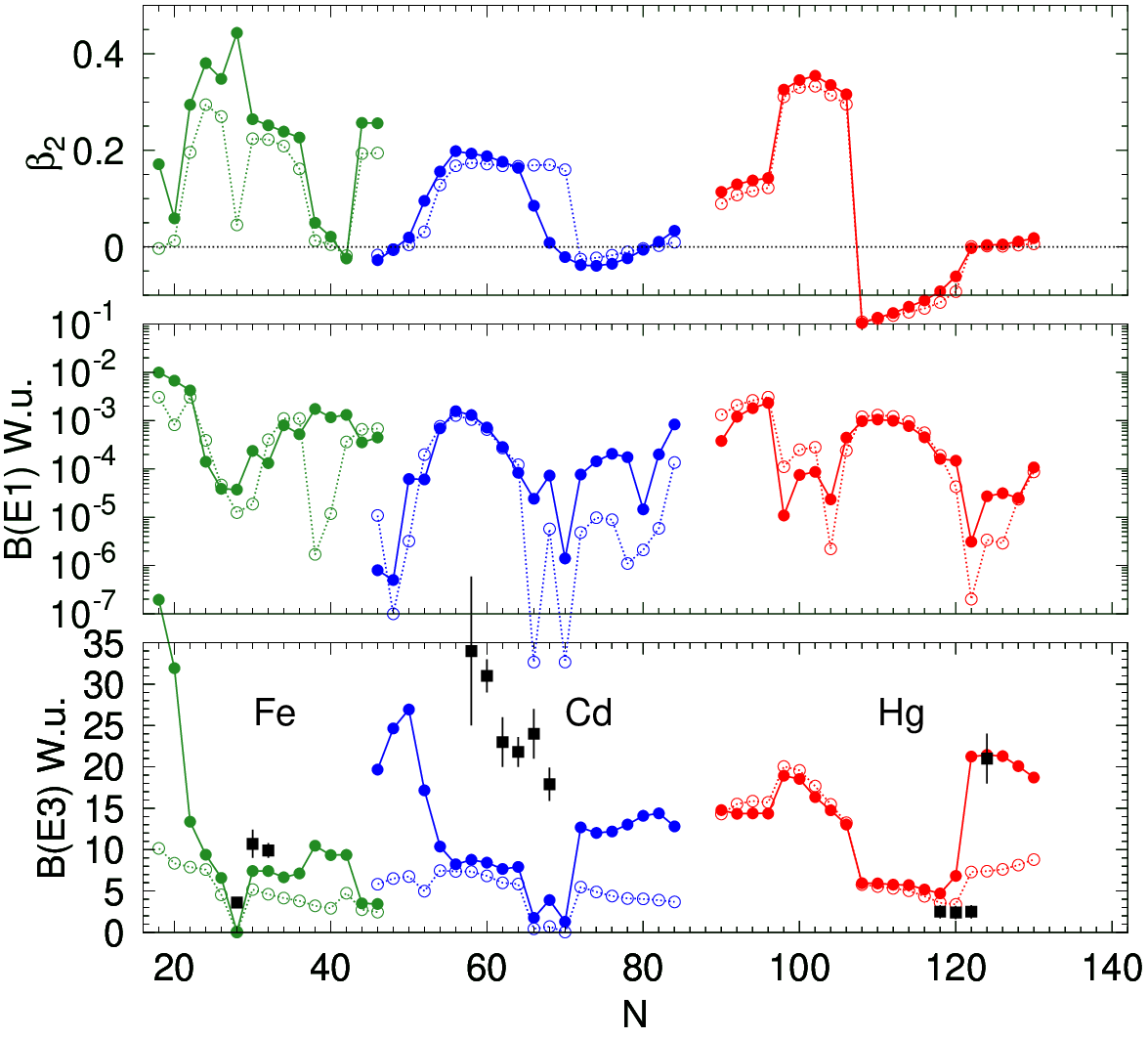}
}
\caption{Same as Fig \ref{fig:1} but for the Fe (Z=26), Cd (Z=48) and Hg (Z=80) isotopic
chains.}
\label{fig:2}       
\end{figure}

Finally, in the middle panel of Fig \ref{fig:1} the E1 strengths are given
as computed with both the rotational formula (open symbols) and the AM
projected wave functions (full symbols). Clearly, the E1 transition is
less collective than the corresponding E3 one, with transitions in the
range of $10^{-3}$ W.u. As a consequence of the reduced collectivity, the
E1 strengths depend more strongly on the details of the single particle structure around
the Fermi level (like which orbital is filled ) than in the deformation 
of the ground and excited states. Therefore, no general trend is observed
as in the E3 case and the E1 strengths are in some cases enhanced and in
some cases reduced by factors as large as two orders of magnitude. Surprisingly,
some dips observed in the evolution of the E1 with neutron number in the 
three isotopic chains when computed with the rotational formula are shifted when 
AM projected wave functions are used. This is the case, for instance in $^{200}$Pb,
$^{192}$Pb or $^{12}$Sn.

In Fig \ref{fig:2} the results corresponding to the isotopic chains with
Z values two units less than magic and corresponding to  Fe ($Z=26$), 
Cd ($Z=48$) and Hg ($Z=80$)
are shown. Contrary to the results of Fig \ref{fig:1} only those isotopes
with neutron number close to a magic number remain spherical in both
their ground and excited states. In those cases, the E3 strength computed
with the AM projected wave functions are enhanced with respect to the 
rotational formula. As neutron number decreases there are transitions
to oblate (Hg) and prolate (Cd and Fe) deformations. The AM projected E3 
strength gets severely quenched and this offers a natural explanation
of the sudden drop in E3 strength in the Hg isotopes. On the other hand,
the comparison with experimental data in the Cd isotopes reveal a serious
mismatch that can be attributed to the fact that, as we are not considering
the quadrupole degree of freedom as dynamical variable, the effects
of shape coexistence present in these isotopes (and observed in our mean
field calculations with the Gogny force) is not accounted for in
the present framework. In the Fe case, the sudden drop in E3 in the isotope
$^{54}$Fe is well reproduced and attributed to the different deformations
of the ground state (spherical) and negative parity excited state (deformed
with $\beta_2=0.45$. Regarding the E1 strengths, in the present case the
differences between the strengths computed with the rotational formula
and the ones with the AM projected wave functions are not so marked as
in the previous case except in the Cd isotopes with $N>62$ and the Fe
isotopes around $N=40$.

%
\begin{figure}
\resizebox{\columnwidth}{!}{%
  \includegraphics{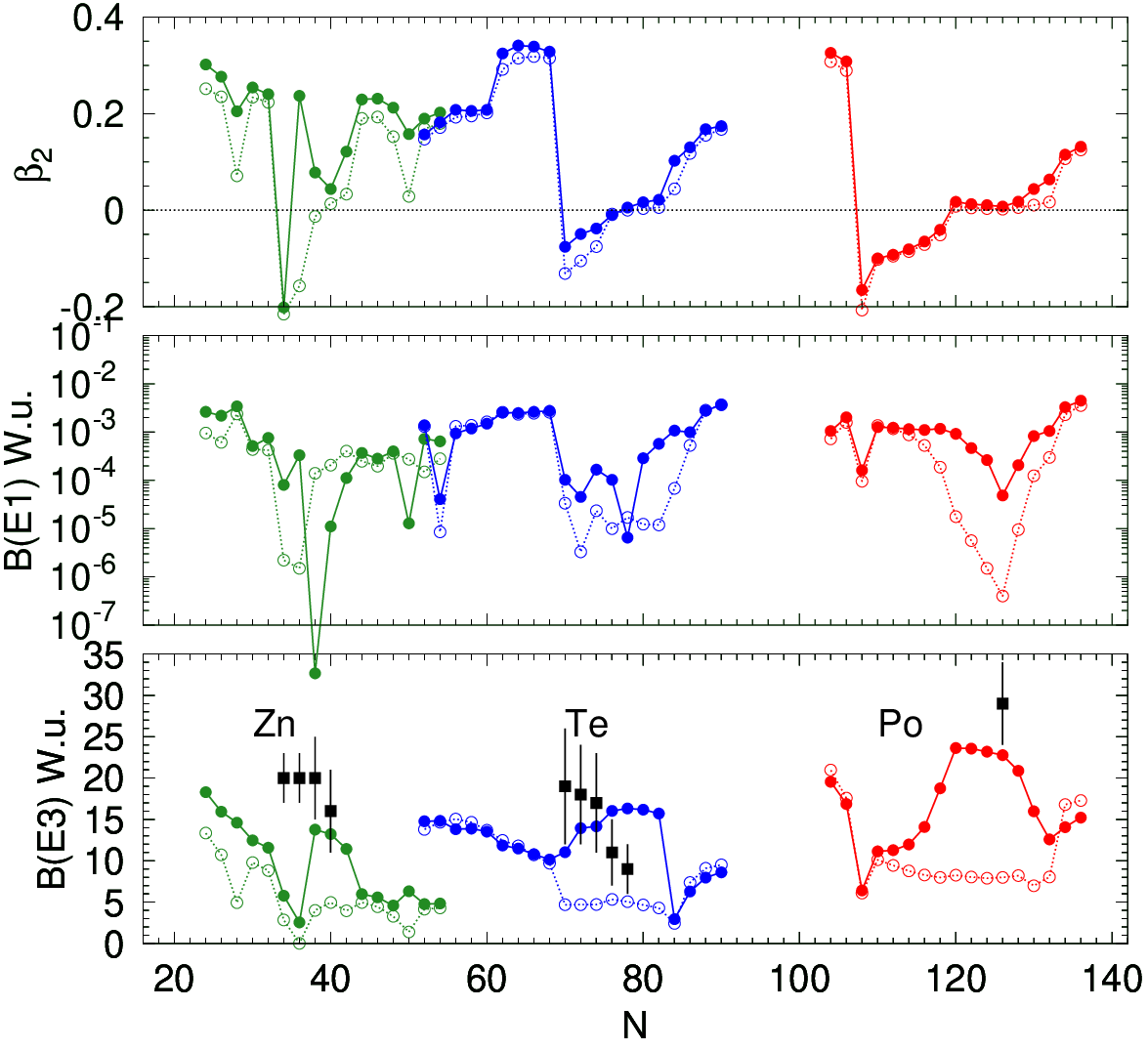}
}
\caption{Same as Fig \ref{fig:1} but for the Zn ($Z=30$), Te ($Z=52$) and 
Po ($Z=84$) isotopic
chains.}
\label{fig:3}       
\end{figure}

In Fig \ref{fig:3} the results corresponding to the isotopic chains with
$Z$ values two units more than magic and corresponding to  Zn ($Z=30$), 
Te ($Z=52$) and Po ($Z=84$) are shown. The results are very similar to the ones
in Fig \ref{fig:2} and corresponding to Fe, Cd and Hg. However, there
are some quantitative differences, as for instance, the broader interval 
of sphericity around magic neutron number in Po, the sudden oblate-prolate
transition in Te or the singular case of $^{64}$Zn with both states oblate.
The latter case was studied in detail in \cite{2014EPJWC..6602091R} where
the relevant role of the simultaneous treatment of quadrupole and octupole
degrees of freedom as a consequence of shape coexistence was discussed.
The same type of coupling could dramatically change the results obtained
in $^{66}$Zn and explaining the severe disagreement with experimental data.
Apart from those two cases, the agreement with experiment is good and, as
in previous cases, the use of AM projected wave functions proves to be
of paramount relevance to obtain sound theoretical predictions. In the
Po case, the E1 strengths computed with the AM projected wave functions
are strongly enhanced with respect to the ones computed with the rotational
formula for those isotopes which are spherical. This is also the case, although
less prominent, for some heavy Te isotopes.

%
\begin{figure}
  \includegraphics[width=\columnwidth]{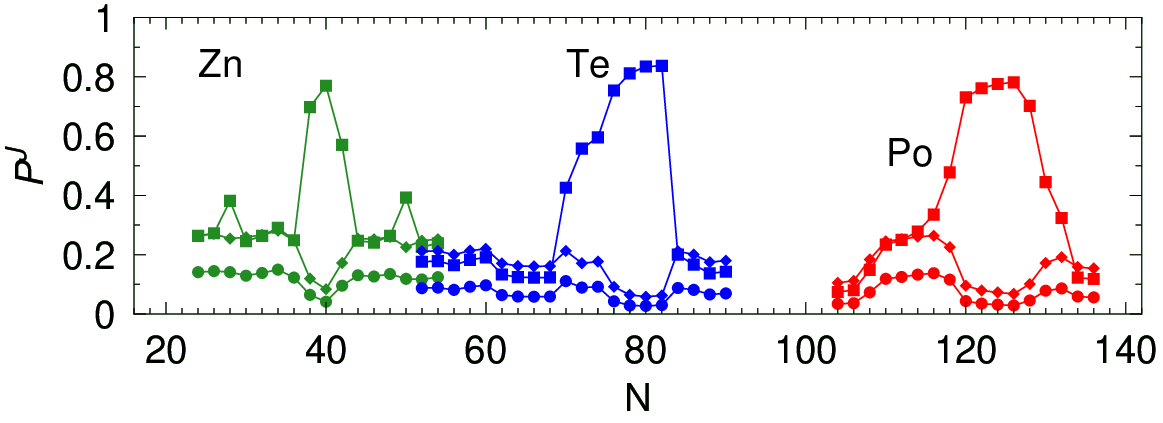}%
  
  \includegraphics[width=\columnwidth]{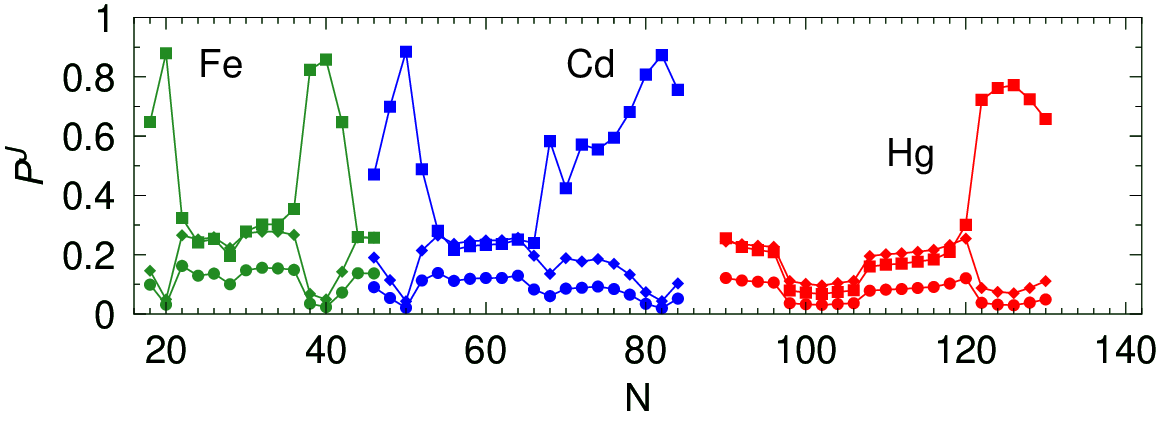}%
  
  \includegraphics[width=\columnwidth]{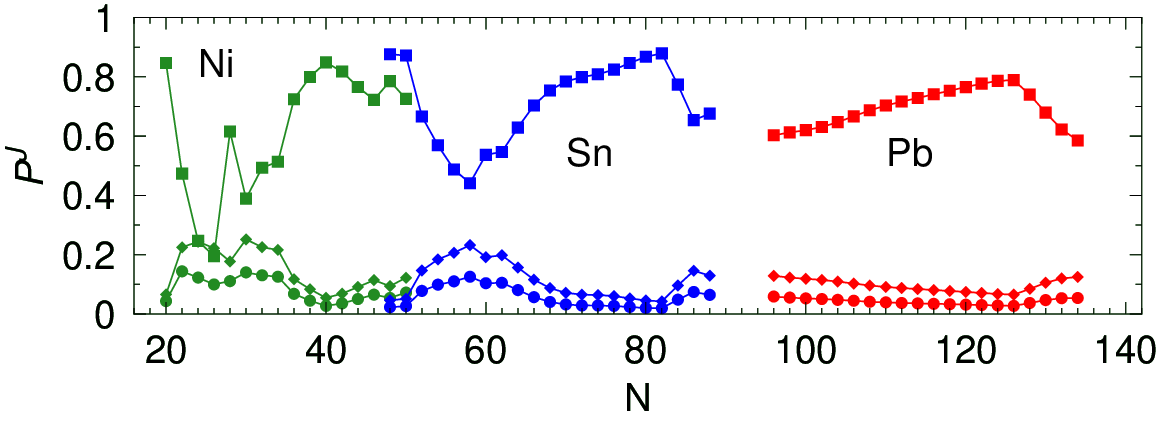}%
\caption{Probability distributions $\mathcal{P}^{J,\pi}$ corresponding to
$J^\pi=1^-$ (bullets), $J^\pi=3^-$ (filled boxes) and $J^\pi=5^-$ (filled diamonds)
for the lowest lying negative parity GCM state of each isotope are plotted
as a function of the neutron number $N$ of the isotope.} 
\label{fig:4}       
\end{figure}

In Fig \ref{fig:4} the probability distribution $\mathcal{P}^{J,\pi}$ of Eq \ref{eq:PJ}
corresponding to the first negative parity intrinsic GCM state is plotted as a function 
of the neutron number for all the nuclei considered
in the present calculations and for $J^\pi$ values $1^-$, $3^-$ and $5^-$.
The results belong to two categories: for spherical nuclei, the probability
of the $3^-$ is very large, typically exceeding 60\%  whereas for deformed
nuclei, and irrespective of the sign of the $\beta_2$ parameter, the probability
distributions for all $J^\pi$ values considered are small, of the order of 10\%,
and very similar for the different $J$ values. In those cases where $\mathcal{P}^{3,-}$
is large we can talk about a vibrational state on top of an spherical ground
state with a physical state in the laboratory system with quantum numbers $3^-$.
This is the case in most of the Ni and all the Sn and Pb isotopes considered.
This feature explains very nicely the strong E3 strength observed, for
instance, in $^{208}$Pb (34 Wu). 
On the other hand, those states with small $\mathcal{P}^{J,\pi}$ values it is
difficult to anticipate which would be the $J^\pi$ of the lowest physical states
and a full AM projected calculation is required to determine the $f_\sigma (Q_3)$
amplitudes of the GCM intrinsic states. 

\section{Conclusions}

In this paper, the range of validity of the rotational formula commonly 
used to evaluate electromagnetic transition strengths is studied in the 
particular case of the E1 and E3 transitions between negative parity 
excited states and the ground state. It is shown that for those cases 
where one or the two quantum mechanic states involved is spherical or 
near spherical the rotational formula breaks down and the use of 
angular momentum projected wave functions is required for a sound 
evaluation of the required overlaps and transition strengths. Several 
isotopic chains with $Z$ values near or at magic values have been 
analyzed and the results compared to experimental data. The E3 
transition strengths consistently show an enhancement of a factor (with 
respect to the rotational formula) between 2 to 4 when the AM projected  
wave functions are used in spherical or nearly spherical nuclei. The 
enhancement proves to be crucial to improve the agreement with 
experimental data. Clearly, such an enhancement is not observed in 
those nuclei with quadrupole deformed configurations where the 
rotational formula remains valid. On the other hand, the E1 transitions 
do not show a clear pattern as in the E3 case. This is due to the much 
less collective character of this transition. The present results 
indicate that for weakly deformed nuclear systems, the use of the 
rotational approximation to compute transition strengths is not 
justified and a more elaborated scheme including angular momentum 
projected wave functions is required. The rotational formula is often
used the other way around, that is to extract $\beta_3$ values from
transition strengths. Obviously this correspondence is wrong for spherical
and near spherical nuclei and its consequences should be handled with care.

\section*{Acknowledgments}

This work was supported by the Ministerio de Econom\'ia y 
Competitividad (MINECO) Spain, under Contracts Nos. FIS2012-34479, 
FPA2015-65929 and FIS2015-63770.

%
\bibliographystyle{epj}
\bibliography{GognyTribute}

\end{document}